\documentclass[runningheads,a4paper]{llncs}
\usepackage{amssymb}
\usepackage{graphicx}
\usepackage{amsmath}
\usepackage{paralist}
\usepackage{mathrsfs}
\usepackage{url}
\usepackage{algorithm}    
\usepackage{algorithmicx}  
\usepackage{algpseudocode}
\usepackage{multirow}  
\usepackage{subfigure}
\usepackage{pgfplots}
\usepackage{makecell}

\usepackage[figuresright]{rotating}
\urldef{\mailsa}\path|{alfred.hofmann, ursula.barth, ingrid.haas, frank.holzwarth,|
	\urldef{\mailsb}\path|anna.kramer, leonie.kunz, christine.reiss, nicole.sator,|
	\urldef{\mailsc}\path|erika.siebert-cole, peter.strasser, lncs}@springer.com|    

\begin{document}
	
		\frontmatter          
	\pagestyle{headings}  
	\addtocmark{An Embedding-based Framework for Recommanding SPARQL Queries} 
	%
	
	%
	%
	\mainmatter              
	\title{TrQuery: An Embedding-based Framework for Recommanding SPARQL Queries}
	\titlerunning{TrQuery: A Framework for Recommanding SPARQL Queries}  
	%
	\author{Lijing Zhang\inst{1,3} \and Xiaowang Zhang\inst{1,3} \and Zhiyong Feng\inst{2,3}}
	\authorrunning{Lijing.} 
	%
	\tocauthor{Zhang et al.}
	\institute{School of Computer Science and Technology, Tianjin University, Tianjin, China,\\
		\and
		School of Computer Software,Tianjin University, Tianjin, China,\\
		\and
		Tianjin Key Laboratory of Cognitive Computing and Application, Tianjin, China\\
	}

	\maketitle
	
\begin{abstract}
	In this paper, we present an embedding-based framework (\texttt{TrQuery}) for recommending solutions of a SPARQL query, including approximate solutions when exact querying solutions are not available due to incompleteness or inconsistencies of real-world RDF data. Within this framework, embedding is applied to score solutions together with edit distance so that we could obtain more fine-grained recommendations than those recommendations via edit distance. For instance, graphs of two querying solutions with a similar structure can be distinguished in our proposed framework while the edit distance depending on structural difference becomes unable. To this end, we propose a novel score model built on vector space generated in embedding system to compute the similarity between an approximate subgraph matching and a whole graph matching. Finally, we evaluate our approach on large RDF datasets DBpedia and YAGO, and experimental results show that \texttt{TrQuery} exhibits an excellent behavior in terms of both effectiveness and efficiency. 
\end{abstract}

\section{Introduction}\label{sec:introduction}
\emph{Resource Description Framework} (RDF), recommended by W3C~\cite{rdf}, is used to represent conceptual description or modeling of information that is implemented in web resources. As the standard query language for RDF graphs (i.e., RDF data),  SPARQL~\cite{sparql} has developed the latest version SPARQL~1.1 \cite{sparql1.1}, which is based on \emph{triple patterns}. The semantics of SPARQL queries are defined as a set of \emph{mappings} (i.e., \emph{solutions}) from triple patterns to RDF graphs via graph matching~\cite{perez_sparql_tods}. There has emerged several SPARQL query engines for evaluation, such as Jena~\footnote{\url{https://jena.apache.org/}}. However, there is not always a (exact) solution of a given SPARQL query evaluating over an RDF graph due to noise or incomplete data in many applications~\cite{incomplete_data}.

\begin{figure*}[h]
	\centering
	\includegraphics[scale=0.36]{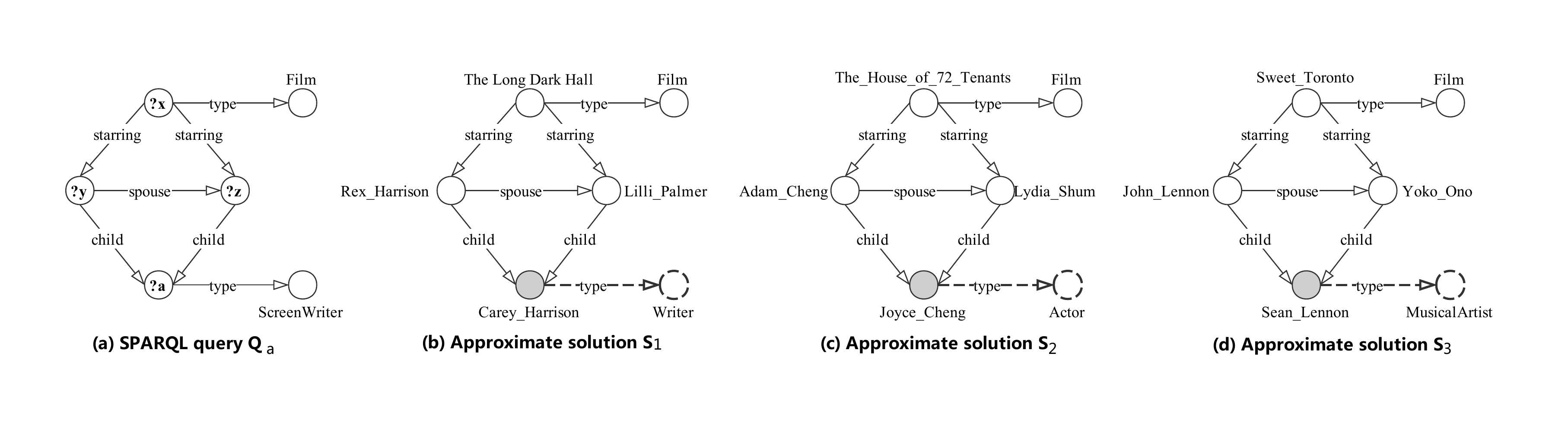}
	\vspace*{-1.0cm}
	\caption{Query graph of $Q_{a}$}\label{fig:query_graph}
	\vspace*{-0.4cm}
\end{figure*}

As a popular treatment, \emph{approximate evaluation} still returns inexact mappings as approximate solutions as similar as possible~\cite{sapper}. 
Recently, there are some approaches to approximately evaluate SPARQL queries~\cite{path alignment,tale,sigma,nema,pbsm}. A key problem of approximate evaluation is to rank inexact mappings of a query with the similarity priority~\cite{pbsm}. However, this problem is challenging since the similarity of graphs is not easy to be quantified. There are some distances proposed to characterize similarity measures~\cite{distance1,distance2,distance3,survey of GED}. The \emph{edit distance}, as a popular similarity measure widely used in many applications~\cite{survey of GED}, provides some edit operations (i.e. the deletion, insertion and substitution of nodes and edges) that is needed to transform one graph to another. Since the edit distance mainly concerns the structural similarity, it can hardly capture the semantic similarity. Thus, the approximate query based on the edit distance is not able to capture good quality matches. Consider the following example.

\noindent {\bf Example 1.} A user wants to find a film that has a couple in the performance, and the child of this couple is a screenwriter. The answer can be obtained by the following SPARQL query, namely $Q_a$, and the query graph is shown in Fig.\ref{fig:query_graph}(a). 
\vspace*{-0.4cm}
\begin{center}
	\fbox{
		\parbox{8.5cm}{
			{\small PREFIX dbo: $<$http://dbpedia.org/ontology/$>$\\
				PREFIX rdf: $<$http://www.w3.org/1999/02/22-rdf-syntax-ns$\#$$>$\\
				SELECT DISTINCT ?film ?actor1 ?actor2 \\
				WHERE$\{$\\
				\indent  \hspace{1.0cm} ?film dbo:starring ?actor1.\\
				\indent  \hspace{1.0cm} ?film dbo:starring ?actor2.\\
				\indent  \hspace{1.0cm} ?actor1 dbo:spouse ?actor2.\\
				\indent  \hspace{1.0cm} ?film rdf:type dbo:film.\\
				\indent  \hspace{1.0cm} ?actor1dbo:child ?child.\\
				\indent  \hspace{1.0cm} ?actor2 dbo:child ?child.\\
				\indent  \hspace{1.0cm} ?child rdf:type dbo:ScreenWriter.\\
				$\}$\\
				\vspace*{-0.3cm}
			}
		}
	}
\end{center}

Unfortunately, there is no exact solutions for this query over DBpedia. The goal of approximate recommendation is that the user can still come up with some reasonable mappings as shown in Fig.~\ref{fig:query_graph}(b)-(d). All of these three matches can be converted to exact matches by only one node substitution operation, thus they will have the same score based on edit distance (score = 1). However, it is clear that the approximate solution $S_1$ is more likely to be an exact mapping, since we can find that the writer \textit{Carey\_Harrison} is the author of 40 stage plays from the Wikipedia. From this example, we can observe the importance of latent semantic information for the approximate query recommendation task.

To this end, in this paper, we propose a novel \emph{embedding-based} framework \texttt{TrQuery} for obtaining more fine-grained recommendations of SPARQL approximate queries, which employs embedding together with the edit distance to compute the score of inexact mappings and return the ranked approximate solution set. 
Given an RDF graph $G$ and a query $Q$, firstly, we embeds entities and relations into \emph{continuous vector spaces} as their features by employing major knowledge graph embedding models, where the inherent structure and semantic of the original RDF data is preserved as much as possible~\cite{representation review}. Secondly, we design a query parser to generate subquery trees that are overlapped parts of all subqueries of $Q$, which can reduce repeated queries to improve efficiency. Then, we define a score model for inexact mappings, which consists of two parts. One part is to score the subquery trees based on statistics of the original RDF graph. The other is built on the vector space to compute the semantic similarity between all the approximate subgraph matchings and the exact subgraph matching. These two parts work together to get the score of the recommended approximate solutions. Extensive experiments were conducted based on the two real-world datasets, i.e., DBpedia~\cite{DBpedia} and YAGO2~\cite{yago2}. The experimental results show that the score model proposed by \texttt{TrQuery} exhibit reasonable rankings.
Furthermover, \texttt{TrQuery} performs a better evaluation of the approximate solutions in terms of both effectiveness and efficiency than the state-of-the-art approximate subgraph matching system SAPPER~\cite{sapper}.

This paper is further organized as follows. In the next section,
we recall background knowledge such as RDF, SPARQL, and embedding.
Section~\ref{sec:framework} introduces the framework of \texttt{TrQuery}. Section~\ref{sec:parser} introduces query parser and Section~\ref{sec:score} introduces recommendation model. Section~\ref{sec:exp} discusses the evaluations of \texttt{TrQuery} and Section~\ref{sec:related} discusses related works. Finally, Section~\ref{sec:con} concludes our works.

\section{Preliminaries}
In this section, we briefly recall some definitions and notations for RDF, SPARQL, and embedding in~\cite{perez_sparql_tods,TransE,TransH,TransR}.

An RDF dataset $G = \{t \mid t \in S \times P \times O\}$ is a set of triples that can be modeled as a labeled directed graph $ G = (V,E,\Sigma,l)$, where $V$ is a finite set of vertices that represent resources, $E \subseteq V \times V$ is a finite set of edges that represent semantic relationships between the resources, and $\Sigma = \mathcal{E} \cup \mathcal{R}$ is a set of labels. The labeling function $ l: V \cup E \rightarrow \Sigma$ maps each vertex or edge to a label in $\Sigma$. Formally, $S = \{ s \mid s = l(v), v \in V, \exists(v,u) \in E \}$, $P = \{ p \mid p = l(\langle v_i,v_j \rangle),\langle v_i,v_j\rangle \in E\}$, $ O = \{ o \mid o = l(v), v \in V, \exists \langle u,v \rangle \in E\}$.

A common SPARQL query contains a group of {\it Basic Graph Pattern} (BGP) queries, whose conjunctive fragment allows to express the core ``{\it Select}$\mid${\it Project}$\mid${\it Join}'' database queries. A series of BGPs can be modeled as a directed labeled graph $Q = (V', E', \Sigma', l')$ where 
\begin{compactitem}
	\item $V'$ is a finite set of vertices;
	\item $E' \subseteq V' \times V'$ is a finite set of edges;
	\item $\Sigma' = \mathcal{E} \cup  \mathcal{R} \cup var$ is a label set where $var$ is a set of variables;
	\item $l': V' \cup E' \rightarrow \Sigma'$ is a labeling function mapping each vertex or edge to a label in $\Sigma'$. 
\end{compactitem}

A solution to a SPARQL query $Q$ over an RDF graph $G$ is a subgraph $G'$ of $G$ for which there exists a function $\varphi$ that maps $var(Q)$ (the variables in $Q$) to either URIs or literals such that $G' = \varphi(Q)$. An approximate solution is a subgraph $G'_a$ of $G$ for which there exists a sequence of edit operations $\phi$, including node and edge insertions, node and edge deletions, and labeling modifications of both nodes and edges, such that $\phi(G'_a) = \varphi(Q)$. 

The \textit{embedding} technique in representation learning is to embed the entities and predicates of the given RDF dataset into continuous vector spaces so as to simplify the manipulation while preserving the inherent structure of the KG, An embedding model $\mathcal{M}$ is a function from an RDF graph $G$ to a vector space $S$, where for each triple $(h, r, t)$ in $G$, $h, t$ are mapped to ${\bf h}$, ${\bf t}$ and ${\bf r}$ in $S$. An embedding model $\mathcal{M}$ defines a score function \texttt{f} and employs $\texttt{f}({\bf h}, {\bf r}, {\bf t})$ to measure its plausibility. Furthermore, a loss function $\mathcal{L}$ is defined to train embedding models. 

Various embedding models have been proposed, which include translation based models and semantic matching models~\cite{embedding_survey}. The former models measure the plausibility of a fact as the distance between the two entities after a translation carried out by the relation, such as \texttt{TransE}~\cite{TransE}, \texttt{TransH}~\cite{TransH}, and \texttt{TransR}~\cite{TransR}. The latter models measure plausibility of facts by matching latent semantics of entities and relations embodied in their vector space representations, such as \texttt{RESCAL}~\cite{rescal}, \texttt{DistMult}~\cite{distmult}, and \texttt{HolE}~\cite{hole}. 
Since our approximate query solution recommendation requires a faster response time, we adopt translation based embeddings. In particular, we employ the state-of-the-art TransE, TransH, and TransR systems to construct embeddings. 

\begin{compactitem}
	\item The score function of TransE is: $g({\bf h}, {\bf r}, {\bf t}) = \left \| {\bf h} + {\bf r} - {\bf t} \right \| _{1/2}$.
	\item The score function of TransH is: ${\bf h}_\perp  = {\bf h} - {\bf w}_r^{\top} {\bf h} {\bf w}_r$ and ${\bf t}_\perp  = {\bf t} - {\bf w}_r^{\top} {\bf t} {\bf w}_r$; $g({\bf h}_\perp, {\bf r}, {\bf t}_\perp) = \left \| {\bf h}_\perp +{\bf r} - {\bf t}_\perp \right \| _{1/2}$ where ${\bf w}$ is a vector projecting entity vectors onto the relation hyperplane.
	\item The score function of TransR is: ${\bf h}_\perp  = {\bf M}_r {\bf h}$; ${\bf t}_\perp  = {\bf M}_r {\bf t}$; and $g({\bf h}_\perp, {\bf t}, {\bf t}_\perp) = \left \| {\bf h}_\perp + {\bf t} - {\bf t}_\perp \right \| _{1/2}$ where ${\bf M}_r$ is a matrix to project entity vectors into a relation-specific space.
\end{compactitem}

\section{The Overview of TrQuery}\label{sec:framework}
In this section, we introduce the overview of \texttt{TrQuery} framework in detail, which is shown in Fig.~\ref{fig:framework}. In particular, \texttt{TrQuery} contains mainly four modules, namely, \emph{Query Parser}, \emph{Embedding Processor},  \emph{Query Executor}, and \emph{Recommendation System}, which are illustrated as follows:
\begin{figure*}[h]
	\centering
	\includegraphics[scale=0.47]{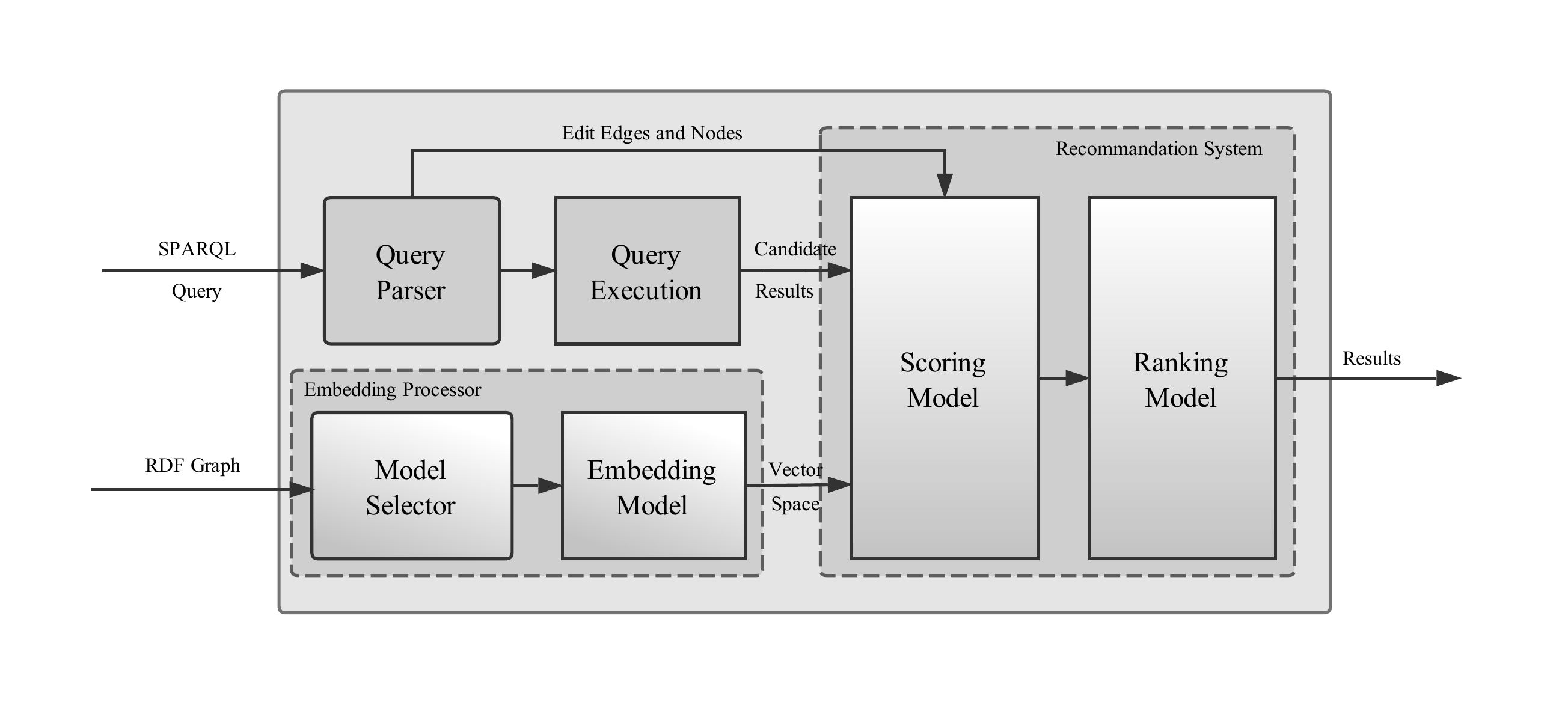}
	\vspace*{-0.6cm}
	\caption{The framework of {TrQuery}}\label{fig:framework}
\end{figure*}

\subsubsection{Query Parser}This module generates a series of subquery trees of a given SPARQL query for reducing duplicate queries and preserving mapping domain, and is detailedly described in Section~\ref{sec:parser}.

\subsubsection{Embedding Processor}This component translates entities and relations to vectors or matrices using embedding techniques in representation learning. It allows a user to select a model from a list of existing embedding models, such as {TransE}, {TransH}, {TransR}, {TransD}~\cite{TransD} etc. 

\subsubsection{Query Executor}This module contains two main parts, namely, SPARQL API and SPARQL query engine, and aims to return mappings of subquery trees as candidate approximate solutions by applying off-the-shelf SPARQL query engine via SPARQL API.

\subsubsection{Recommendation System}This module aims to score mappings of candidate approximate queries for recommending approximate solutions based on our proposed score models which are defined in detail in Section~\ref{sec:score}.

\section{The Query Parser of TrQuery}\label{sec:parser}

In this section, we present a query parser of \texttt{TrQuery}, namely \texttt{TrQuery-QP}, to generate subquery trees for a given query, which can retain the mapping domain and reduce duplicate queries. Here, \emph{mapping domain} indicates the variables in the BGPs of the given SPARQL query. Retaining the mapping domain is to enable users to get a complete solution. In example 1, the user wants to get whole matches of (?film, ?actor1, ?actor2, ?child), therefore the matches of sub-domain, such as (?film, ?actor1), is meaningless and worthless.

Given a threshold $t$ of edit distance and a SPARQL query $Q$, it is a straightforward way to evaluate exact matches of the query graphs whose edit distance are less than $t$ generated from $Q$. In this method, the approximate query evaluation can be transformed as a series of exact query problems. However, there may potentially produce many subquery graphs with edit distance less than $t$. In addition, we can observe that there are several overlapping parts among these query graphs. Therefore, it is beneficial to query the overlapping parts first since they could be used duplicately. 

\renewcommand{\algorithmicrequire}{ \textbf{Input:}}
\renewcommand{\algorithmicensure}{ \textbf{Output:}}
\begin{algorithm}[H]
	\caption{TrQuery-QP}\label{alg1}
	\begin{algorithmic}[1]
		\Require Query graph $Q = (V,E)$
		\Ensure  A set of subquery trees $Q_{t}S$
		
		\Function{delConstantLeaf}{$Q$}
		\State $R_l \gets$ all leaf nodes in $Q$;
		\State $R_{l_s} \gets$ the nodes attached constants in $R_l$;
		\While {$R_{l_s} \neq empty$}
		\State
		$R_{C} \gets$ the edges connected by the nodes in $R_{l_s}$;
		\State 
		$R_{N} \gets$ the nodes in $R_{l_s}$;
		\State $V \gets V-R_{N}$, $E \gets E-R_{C}$, $Q_{new} \gets (V, E)$;
		\State $R_l \gets$ all leaf nodes in $Q_{new}$;
		\State $R_{l_s} \gets$ the nodes attached constants in $R_l$;
		\EndWhile
		\State \Return $Q_{new}$
		\EndFunction
		\State $Q_{new} \gets delConstantLeaf(Q)$;
		\State $E_{tree} \gets combntns(Q_{new}.E,|Q_{new}.V|-1)$;
		\For {$E_{t}$ in $E_{tree}$}
		\State $Q_{tree} \gets (V, E_{t})$;
		\If{$Q_{tree}$ is connected}
		\State $Q_{tree} \gets delConstantLeaf(Q_{tree}) $;
		\State $Q_{t}S \gets $ append $Q_{tree}$ to $Q_{t}S$;
		\EndIf
		\EndFor 
		\State \Return $Q_{t}S$
	\end{algorithmic}
\end{algorithm}

We describe the steps of \emph{Query Parser} of \texttt{TrQuery} in Algorithm~\ref{alg1}. Given a query graph $Q$, we treat it as an undirected graph. We design a function $DelConstantLeaf$ to remove the constant (URI or literal) attached with the leaf nodes (degree = 1) and remove the edges that are connected to these nodes. The effect of this function is to retain the mapping domain, and relax the query conditions as much as possible. Firstly, in line 12, $DelConstantLeaf$ is invoked to get a new query graph named $Q_{new}$ in which all leaf nodes are attached with variables. 
In line 13, the function $combntns$ is used to generates all combination of $|Q_{new}.V|-1$ edges from $Q_{new}.E$, which has a total of $C_{|Q_{new}.E|}^{|Q_{new}.V|-1}$ cases. Here, We define $Q_{new}.E$ as the set of edges in $Q_{new}$. Then, in lines 14-20, we produce all the spanning trees of query graph $Q_{new}$ which are merged into a set named $Q_{t}S$. In line 15, we construct a new query $Q_{tree}$ which contains all nodes and $|Q_{new}.V|-1$ edges in $Q_{new}$. In line 16, we judge whether $Q_{tree}$ is connected, if so, it is the spanning tree of $Q_{new}$. In line 17, $DelConstantLeaf$ is invoked to update $Q_{tree}$, and in line 18 add the query tree to $Q_{t}S$.

\begin{figure}[h]
	\centering
	\includegraphics[scale=0.4]{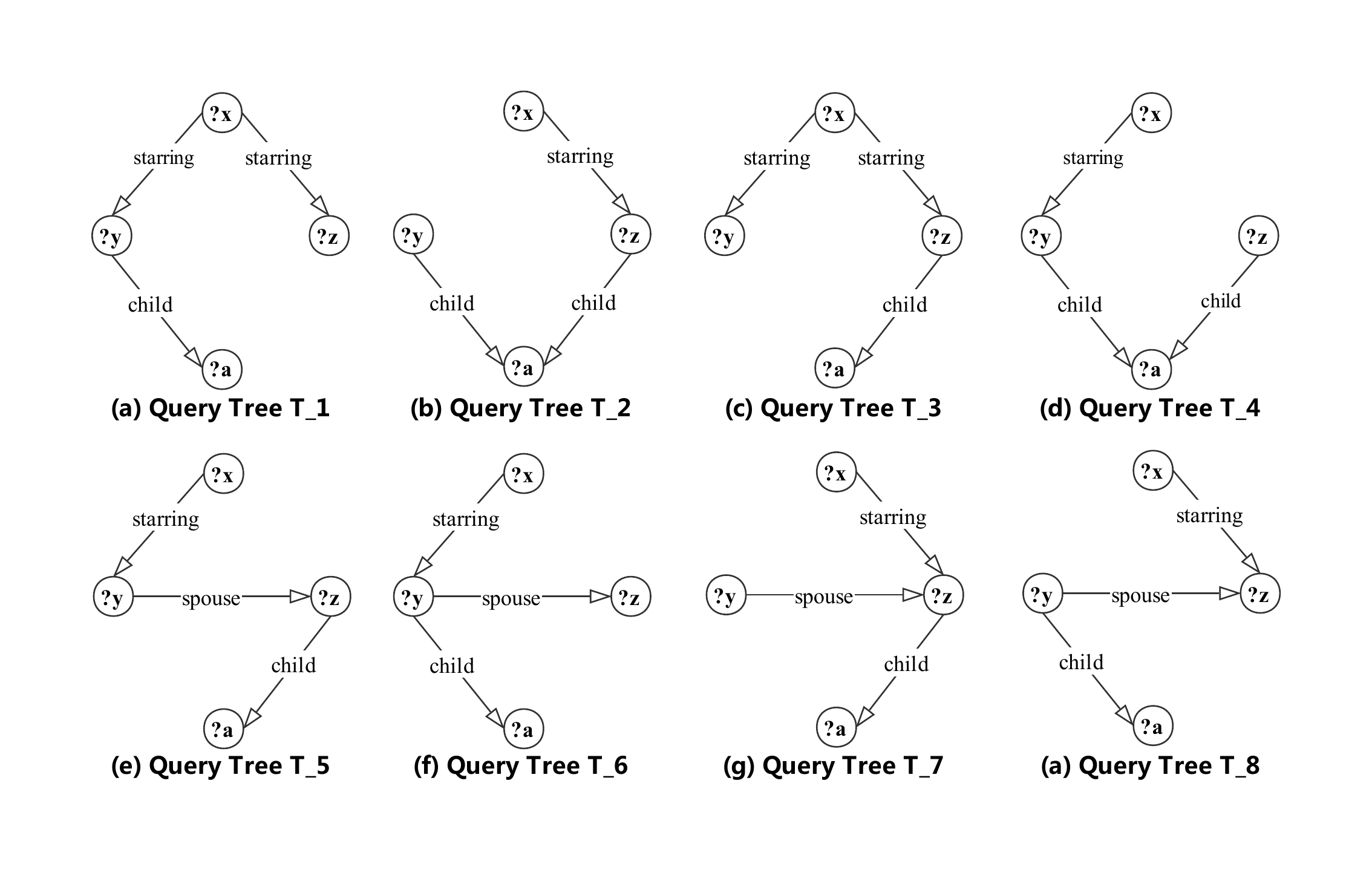}
	\caption{Example for query parser of $Q_a$}\label{fig:query parser example}
\end{figure}

For example, we show the subquery trees of the example query $Q_{a}$ in Section~\ref{sec:introduction} generated by Algorithm~\ref{alg1} in Fig.~\ref{fig:query parser example}. These subquery trees are the overlapping parts of all subqueries of $Q_{a}$, which should be done first to reduce repeated queries. 


\section{The Recommendation System of TrQuery}\label{sec:score}

In this section, we present a recommendation system of \texttt{TrQuery}, namely \\ \texttt{TrQuery-ASR}, to recommend approximate solutions. The recommendation model is composed of the following two aspects: \emph{scoring model} and \emph{ranking model}, which is to measure the goodness of a matching and return a reasonable ranked solution set.

\subsection{Scoring model}

We assume that $Q$ denotes a BGP, $r$ denotes a relation (label of edge), $\mathrm{dom}(r)$ and $\mathrm{ran}(r)$ denote the domain and range of $r$, that is, the set of head and tail entities of $r$ respectively. Let $c$ be a constant (URI or literal). $\mathrm{dom}(r, c)$ and $\mathrm{ran}(c, r)$ denote the domain and range of $r$ restricting at $c$, respectively. 

\begin{definition}[Index of graph]
	Let $Q=(V', E', \Sigma', l')$ be a set of BGPs and $e$ be an edge. $\mathrm{I}(Q)$ is a positive integer defined as follows:
	\[
	\mathrm{I}(Q):= \sum_{e \in E'}\, \delta(e)
	\]
	\begin{compactitem}
		\item $\delta(e) =\frac{|\mathrm{dom}(r)|+|\mathrm{ran}(r)|} {2}$, if $e$ is of the form $(?x, r, ?y)$;
		\item $\delta(e) = |\mathrm{dom}(r, c)|$, if $e$ is of the form $(?x, r, c)$;
		\item $\delta(e) = |\mathrm{ran}(c, r)|$, if $e$ is of the form $(c, r, ?y)$.
	\end{compactitem}
\end{definition}

The $\mathrm{I}(Q)$ represents the amount of information covered by the query graph $Q$ over the RDF graph $G$. 

\begin{definition}[Weight of edge]
	Let $Q=(V', E', \Sigma', l')$ be a set of BGPs and $e$ be an edge. $	\mathrm{w}(e)$ denotes the weight of $e$ defined as follows:
	\[
	\mathrm{w}(Q, e): = \frac{\mathrm{I}(Q)}{\delta(e)}
	\]
\end{definition}

The $\mathrm{weight}$ describes the importance of each edge in the query graph, that is, the larger the weight value, the more restrictive this edge is.

\begin{definition}[Score of graph]
	Let $Q=(V', E', \Sigma', l')$ be a set of BGPs and $e$ be an edge. $	\mathrm{Score}(Q)$ is a positive value defined as follows: 
	\[
	\mathrm{Score}(Q): = \sum_{e \in E'}\, w(Q,e)
	\]
\end{definition}

For {example 1}, the score of the subquery trees in Fig.~\ref{fig:query parser example} can be calculated through the above method. In detail, we get the $\delta(e)$ through SPARQL queries. For instance, $\delta$($\langle$?film,rdf:type,Film$\rangle$) can be get by the following SPARQL query, and the idea is easy to generalize to other edges.

\begin{center}
	\fbox{
		\parbox{8.5cm}{
			{\small PREFIX dbo: $<$http://dbpedia.org/ontology/$>$\\
				PREFIX rdf: $<$http://www.w3.org/1999/02/22-rdf-syntax-ns$\#$$>$\\
				SELECT COUNT(DISTINCT ?film)\\
				WHERE$\{$\\
				\indent  \hspace{1.0cm} ?film rdf:type dbo:film.\\
				$\}$\\
				\vspace*{-0.3cm}
			}
		}
	}
\end{center}

To employ embedding models, we firstly embed types (i.e., classes) which is the important notion of RDF graphs while the current embedding model excludes.
\begin{definition}[Embedding of type]
	Let $G$ be an RDF graph and $t$ be a type. The vector $\mathbf{{t}}$ of a type is defined as follows:
	\[
	\mathbf{{t}}: = \frac{\mathbf{e}_1+ \cdots +\mathbf{e}_m}{m}
	\]
	Here $|\{e_i \mid (e_i, type, t)  \in G\}| = m$ and $\mathbf{e}_1, \ldots, \mathbf{e}_m$ is vectors after embedding.
\end{definition}

\begin{definition}[Extended embedding-based triple score]
	Let $G$ be an RDF graph and $(h, r, t)$ be a triple in $G$. We use $g^{\ast}(h, r, t)$ to denote an \emph{Extended embedding-based triple score} as follows:
	\[
	g^{\ast}(h, r, t):= \left\{
	\begin{aligned}
	\left \| {\bf h} - {\bf t} \right \| _{1/2}, & & r \text{ is rdf:type};\\
	g(h,r,t), & & \text{otherwise}.\\
	\end{aligned}
	\right.
	\]
	
\end{definition}

\begin{definition}[Normalization]\label{D6}
	Let $G$ be an RDF graph and $(h, r, t)$ be a triple in $G$. $g(h,r,t)$ denotes the embedding-based triple score. We use $f(h,r,t)$ denote the \emph{normalization} of $g^{\ast}(h, r, t)$ defined as follows:
	\[
	f(h,r,t)=\left\{
	\begin{aligned}
	1~~~~~~~~, & & (h,r,t) \in G;\\
	\frac{1}{1+g^{\ast}(h,r,t)}, & & (h,r,t) \notin G.\\
	\end{aligned}
	\right.
	\]
\end{definition}

This normalization function guarantees that the score is equal to 1 only if the triple exists in the RDF graph, otherwise it must be less than 1. Moreover, the closer the value of $f(h,r,t)$ is to 1, the more reasonable  the triple is. Here, to determine whether the triple belongs to $G$, we use the \textit{ASK} query in SPARQL to implement it. For example, it will return \textit{false} for the following SPARQL query.

\begin{center}
	\fbox{
		\parbox{8.5cm}{
			{\small PREFIX dbo: $<$http://dbpedia.org/ontology/$>$\\
				PREFIX rdf: $<$http://www.w3.org/1999/02/22-rdf-syntax-ns$\#$$>$\\
				PREFIX dbr: $<$http://dbpedia.org/resource/$>$\\
				ASK$\{$\\
				\indent  \hspace{1.0cm}  dbr:Carey\_Harrison rdf:type dbo:ScreenWriter.\\
				$\}$\\
				\vspace*{-0.3cm}
			}
		}
	}
\end{center}

\begin{definition}[Score of solution]
	Let $G$ be an RDF graph, $Q$ be a query, $\mu$ be a mapping. We use $\mathrm{Score}(G, Q, \mu)$ to denote the score of $\mu$ w.r.t. $Q$ in $G$ defined as follows:
	
	\[
	\mathrm{Score}(G, Q, \mu):= \sum_{e\in Q, (h,r,t) \in \mu(Q)}  \mathrm{w}(Q, e) \ast f(h,r,t)
	\]
	
	Here $\mu(Q)$ is a set of triples by substituting $a$ for $?x$ for all $?x \to a \in \mu$. 
\end{definition}

\begin{proposition}[Exactness Protectability]\label{p1}
	Let $G$ be an RDF graph and $Q$ be a query. For any exact mapping $\mu$ of $Q$ over $G$, $\mathrm{Score}(G, Q, \mu)$ is maximal. This can be readily proved by Definition~\ref{D6}.
\end{proposition}

For Example 1 in Fig.~\ref{fig:query_graph}, there are three approximate solutions for the given query $Q_a$. The difference between these three solutions is the match for $?child$. Here, $Q_a'$ is the remaining part of $Q_a$ after deleting the node labeled by \textit{ScreenWriter} and the edge $\langle$?child, rdf:type, ScreenWriter$\rangle$. Then, 

\begin{compactitem}
	\item $\mathrm{Score}(G, Q_a, S_1) = Score(G,Q_a',S_1) \\+w(Q_a,e) \ast f($\textit{Carey\_Harrison,rdf:type,ScreenWriter}$)$;
	\item $\mathrm{Score}(G, Q_a, S_2) = Score(G,Q_a',S_2) \\+w(Q_a,e) \ast f($\textit{Joyce\_Cheng,rdf:type,ScreenWriter}$)$;
	\item $\mathrm{Score}(G, Q_a, S_3) = Score(G,Q_a',S_3) \\+w(Q_a,e) \ast f(\textit{Sean\_Lennon,rdf:type,ScreenWriter})$.
\end{compactitem}

Since $S_1$,$S_2$ and $S_3$ are exact mappings for $Q_a'$, the $Score(G,Q_a',S_i), (i=1,2,3)$ are equal. Therefore, the factor that determines the ranking result is the \emph{rationality} of the triples \textit{(Carey\_Harrison, type, ScreenWriter)}, \textit{(Joyce\_Cheng, type, ScreenWriter)} and \textit{(Sean\_Lennon, type, ScreenWriter)}.

\subsection{Ranking model}

In order to be able to quickly sort the approximate solutions, we apply the \textit{Timsort algorithm}, which derives from merge sort and insert sort, and has a much smaller time-space complexity than other sorting algorithms. This part will return the Top-K approximate solutions to users.

Next, Algorithm~\ref{alg2} shows how to recommend approximate solutions based on scores of inexact mappings. For each query tree generated by \emph{Query Parser}, we firstly use the \textit{Query Executor} to get the candidate inexact mappings (line 3). Then, for each mapping (line 4), we calculate the edit distance, if the edit distance is less than the given threshold $t$, we calculate the score of this mapping and add it to the approximate solution set $S$ (lines 5-8). Although there may be a large number of approximate solutions for a query, the users are merely interested in the top-K results. We sort the matches based on their scores to obtain the top-K results (line 11). 

\renewcommand{\algorithmicrequire}{ \textbf{Input:}}
\renewcommand{\algorithmicensure}{ \textbf{Output:}}
\begin{algorithm}
	\caption{TrQuery-ASR}\label{alg2}
	\begin{algorithmic}[1]
		\Require RDF $G = (V,E)$, subquery trees set $Q_{t}S$, threshold $t$, an positive integer K
		\Ensure Ranked candidate approximate solutions $S$
		\State $S \gets \emptyset$;
		\For{$i \gets 1:length(Q_{t}S)$}
		\State $C_S \gets SparqlAPI(G,Q_{t}S[i])$;
		\For{$\mu$ in $C_S$}
		\If {$editDistance(\mu(Q_{t}S[i]),Q) < t$}
		\State score = Score(G,Q,$\mu$);
		\State $S \gets$ append $(\mu, score)$ to $S$;
		\EndIf
		\EndFor 
		\EndFor 
		\State $S \gets reverseSort(S,K)$;
		\State \Return $S$
		
	\end{algorithmic}
\end{algorithm}

\section{Experiments and Evaluations}\label{sec:exp}

In this section, we evaluate the performance of our system \texttt{TrQuery}, which is implemented in Python. We conducted extensive experiments to verify the efficiency and scalability of the proposed algorithms on real-world datasets. 

\paragraph{\underline{Dataset}}  We implement \texttt{TrQuery} on two benchmark RDF datasets, DBpedia and YAGO. The number of nodes, edges, triples and types of the two data sets are shown in the following table. 
\begin{center}
	\scalebox{1.0}{
		\begin{tabular}{|l|c|c|c|c|}
			\hline
			{\bf Dataset} & {\bf Entity} & {\bf Relation} & {\bf Triple}  & {\bf Type} \\
			\hline
			DBpedia~ & ~6,099,488~ & ~659~   & ~18,154,761~ & ~14989~ \\
			\hline
			YAGO~  & ~4,295,827~ & ~38~    & ~23,243,143~ & ~4,987~ \\
			\hline
		\end{tabular}%
	}
\end{center}%
\paragraph{\underline{Experiment setup}} 

In our experiments, (1) edit distance threshold $t$ was set as 2, (2) all the experiments were conducted on a server with one 4-Core CPUs (Intel i5 3.10GHz), 20GB RAM, and Ubuntu 14.04 operation system.
\vspace*{-0.1cm}
\subsection{Efficiency Evaluation}

\subsubsection{\textbf{Experiment 1. Comparison with exact SPARQL query engine}}  Neither DBpedia nor YAGO has provided standard queries. In this experiment, we formulated 10 queries in SPARQL of different complexities (i.e. number of nodes, edges, variables and structure of query graphs) that have exact solutions, namely $Q_1$-$Q_{10}$, for each dataset. Then we obtained another ten queries by adding one triple pattern to $Q_1$-$Q_{10}$ such that they have no exact solutions, namely $Q_1'$-$Q_{10}'$. 

\begin{figure*}[h]
	\centering
	\scalebox{0.75}{
		\begin{tikzpicture}
		\begin{semilogyaxis}[
		width=15cm,height=5cm,
		bar width=5pt,
		symbolic x coords={$\mathbf{Q}_1(')$,$\mathbf{Q}_2(')$,$\mathbf{Q}_3(')$,$\mathbf{Q}_4(')$,$\mathbf{Q}_5(')$,$\mathbf{Q}_6(')$,$\mathbf{Q}_7(')$,$\mathbf{Q}_8(')$,$\mathbf{Q}_9(')$,$\mathbf{Q}_{10}(')$},
		x tick label style={rotate=-15,anchor=north},
		ylabel={Runtime(msec)},
		xlabel={},
		ymin=0,ymax=1000000,
		ymajorgrids=true,
		xlabel style={below=0.1cm},
		enlarge x limits=0.05,
		legend style={at={(0.5, -0.23)}, 
			anchor=north,
			legend columns=-1, 
			ylabel near ticks,
			ylabel style={right=0.3cm},
		},
		ybar,
		]
		\addplot
		coordinates {
			($\mathbf{Q}_1(')$,132)($\mathbf{Q}_2(')$,7)($\mathbf{Q}_3(')$,120)($\mathbf{Q}_4(')$,16)($\mathbf{Q}_5(')$,12)($\mathbf{Q}_6(')$,4)($\mathbf{Q}_7(')$,9)($\mathbf{Q}_8(')$,12)($\mathbf{Q}_9(')$,26)($\mathbf{Q}_{10}(')$,110)
		};
		
		\addplot
		coordinates {
			($\mathbf{Q}_1(')$,6267)($\mathbf{Q}_2(')$,1744)($\mathbf{Q}_3(')$,74415)($\mathbf{Q}_4(')$,1246)($\mathbf{Q}_5(')$,924)($\mathbf{Q}_6(')$,172)($\mathbf{Q}_7(')$,215)($\mathbf{Q}_8(')$,769)($\mathbf{Q}_9(')$,849)($\mathbf{Q}_{10}(')$,2748)
		};

		\legend{Exact query for Q, Approximate query for Q'}
		
		\end{semilogyaxis}
		\end{tikzpicture}
	}
	\vspace*{-0.2cm}
	\caption{Comparison of approximate queries and exact queries in runtime }\label{fig:experiment1}
\end{figure*}
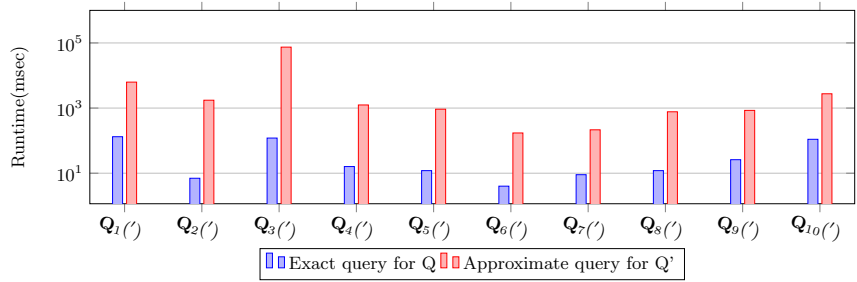

\begin{figure}[!tbp]
	\centering
	\includegraphics[scale=0.5]{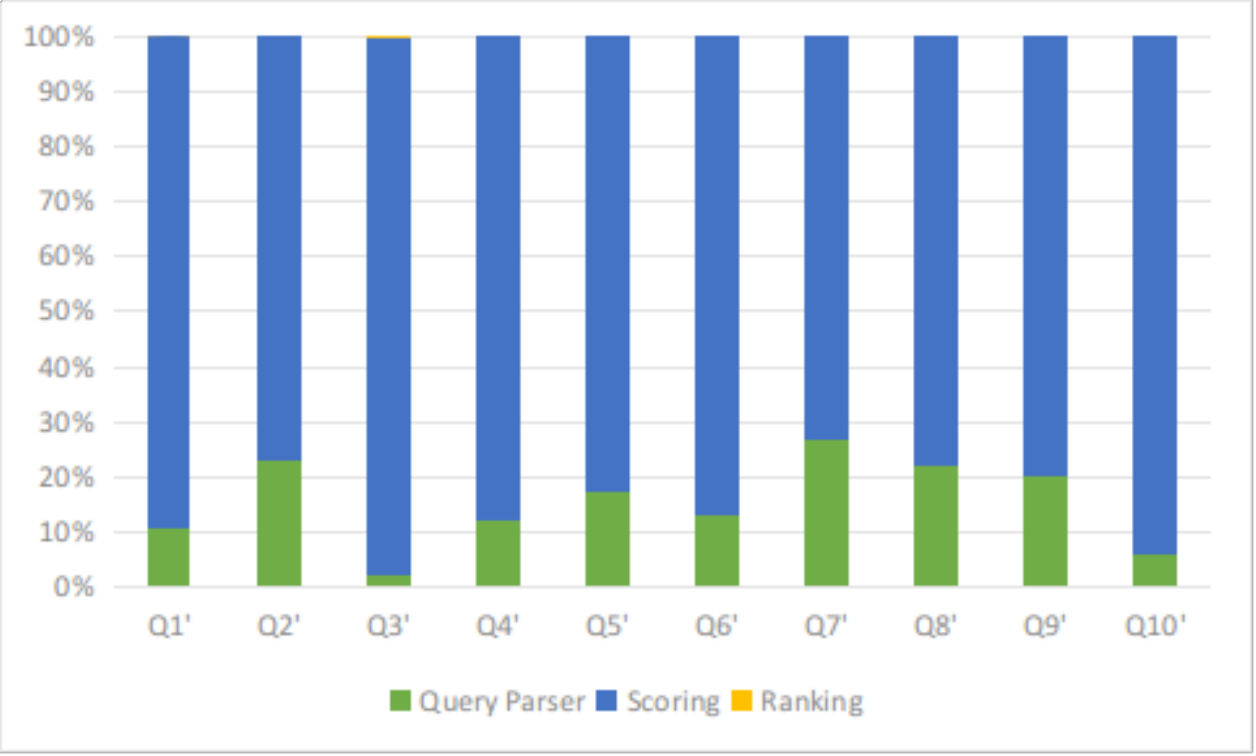}
	\caption{Cumulative time percentage of each step}\label{fig:Cumulative time}
\end{figure}

We evaluate the exact query time of $Q_1$-$Q_{10}$ of Jena and the approximate query time of $Q_1'$-$Q_{10}'$ via \texttt{TrQuery}. We evaluated each query 10 times and measured the average response time in \textit{msec}, including the time of \textit{query parsing}, \textit{scoring} and \textit{ranking}, which are shown in the Fig.~\ref{fig:experiment1}. It is obvious that the time for exact query is shorter than approximate query by \texttt{TrQuery}. Fortunately, \texttt{TrQuery} can recommend approximate solutions within an \emph{acceptable} computation time. Then, we analyze the time percentage of each step in \texttt{TrQuery} as shown in Fig.~\ref{fig:Cumulative time}. For most queries, the most amount of time is spent for the \textit{scoring} step (i.e. 70\% of the cumulative amount of time in average), and next is \textit{query parsing} step (i.e. 28\% of the cumulative amount of time in average) . The reasons are: (i) in order to get all approximate solutions we generate all spanning trees for the query graph, (ii) we calculate scores for all candidate solutions which would be a large set.

\subsubsection{\textbf{Experiment 2. Scalability of \texttt{TrQuery}}}
In this experiment we analysed in more depth for evaluating the scalability of \texttt{TrQuery}. There are three questions to be answered: (1) how the structure of query graph affects the query time; (2) how the number of edges in the query graph affects the query time; (3) how embedding model of \texttt{TrQuery} affects the query time. Here, we calculate the runtime of each answer by dividing the total time by the number of solutions which represents the average time of per solution. To answer the above questions, we have done the following experiments.

\begin{figure}[!tbp]
	\centering
	\subfigure[\scriptsize{The histogram with queries structure}]{\label{fig:a1}
		\begin{minipage}[t]{0.45\linewidth}
			\centering
			\scalebox{0.55}{
				\begin{tikzpicture}
				\begin{axis}[
				width=8cm,height=6cm,
				bar width=7pt,
				symbolic x coords={$\mathbf{Q}_1$,$\mathbf{Q}_2$,$\mathbf{Q}_3$,$\mathbf{Q}_4$,$\mathbf{Q}_5$},
				x tick label style={rotate=-15,anchor=north},
				ylabel={Runtime of each answer (msec)},
				xlabel={Query 1$\sim$5},
				ymin=0,ymax=12,
				ymajorgrids=true,
				xlabel style={below=0.1cm},
				enlarge x limits=0.15,
				legend style={at={(0.75, -0.3)}, 
					legend columns=-1, 
					ylabel near ticks,
					ylabel style={right=0.3cm},
				},
				ybar,
				]
				\addplot
				coordinates {
					($\mathbf{Q}_1$,1.562)($\mathbf{Q}_2$,3.521)($\mathbf{Q}_3$,2.797)($\mathbf{Q}_4$,3.571)($\mathbf{Q}_5$,5.796)
					
				};
				
				\addplot
				coordinates {
					($\mathbf{Q}_1$,2.763)($\mathbf{Q}_2$,5.321)($\mathbf{Q}_3$,4.677)($\mathbf{Q}_4$,5.583)($\mathbf{Q}_5$,9.896)
					
				};
				
				\legend{DBpedia, YAGO}
				
				\end{axis}
				\end{tikzpicture}
			}
		\end{minipage}
	}
	\subfigure[\scriptsize{The trendline with number of edges}]{\label{fig:a}
		\begin{minipage}[t]{0.45\linewidth}
			\centering
			\scalebox{0.55}{
				\begin{tikzpicture}
				\begin{axis}[
				width=8cm,height=6cm,
				symbolic x coords={4,6,8,10,12},
				x tick label style={rotate=-15,anchor=north},
				ylabel={Runtime of each answer (msec)},
				xlabel={number of edges},
				ymin=0,ymax=100,
				ymajorgrids=true,
				grid style=dashed,
				anchor=north,
				legend pos = north west,
				xlabel style={below=0.1cm},
				legend style={at={(0.16, -0.3)},
					legend columns=-1, 
					ylabel near ticks,
					ylabel style={right=0.3cm},
				},
				]

				\addplot[
				color=blue,
				mark=square*, mark options={fill=blue}
				]
				coordinates {
					(4,1.518)(6,19.333)(8,22.212)(10,25.392)(12,85.527)
				};
				
				\addplot[
				color=red,
				mark=triangle*, mark options={fill=blue}
				]
				coordinates {
					(4,3.58)(6,26.333)(8,35.22)(10,48.92)(12,95.57)
				};

				\legend{DBpedia, YAGO}
				
				\end{axis}
				\end{tikzpicture}
			}
		\end{minipage}

	}
	\vspace*{-0.2cm}
	\caption{The effect of query complexity on query time}\label{fig:positive}
\end{figure}
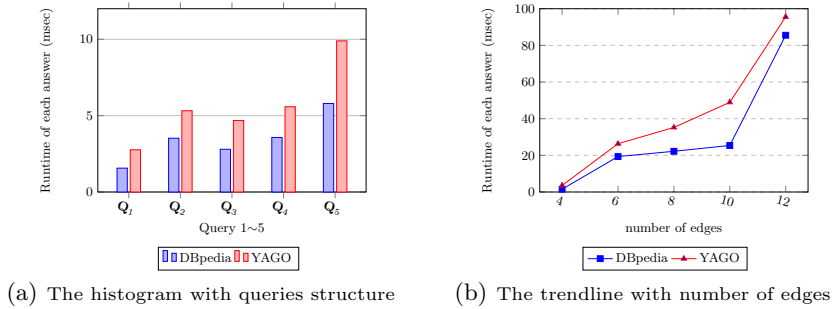

\paragraph{\textbf{Experiment 2.1}} We formulated 5 queries with different structures and the same number of nodes and variables, namely $Q_{11}$-$Q_{15}$, to determine how the structure of the query graph affects the query time. Each of these 5 queries contains 6 nodes with 4 variables, and on the same topic. The shapes of the queries are: \textit{line}, \textit{star}, \textit{ring}, \textit{line\&star}, and \textit{line\&star\&ring}, respectively. The query runtimes are shown in Fig.~\ref{fig:a1}, which indicates that line query consumes the shortest time. By analyzing, the reason is that the query with line structure only has one spanning tree.

\paragraph{\textbf{Experiment 2.2}} We design another 5 queries with the same shape ``\textit{line\&star\&ring}'' and the different number of edges, namely $Q_{16}$-$Q_{20}$. The trendline are shown in Fig.~\ref{fig:a}. It can be clearly seen that the more the number of edges, the more query time is needed. The reason is obviously that the more edges the more spanning trees are generated, the greater candidate set is, and the more time the scoring step takes.

\begin{figure*}[!tbp]
	\centering
	\scalebox{0.75}{
		\begin{tikzpicture}
		\begin{semilogyaxis}[
		width=15cm,height=5cm,
		bar width=5pt,
		symbolic x coords={$\mathbf{Q}_1'$,$\mathbf{Q}_2'$,$\mathbf{Q}_3'$,$\mathbf{Q}_4'$,$\mathbf{Q}_5'$,$\mathbf{Q}_6'$,$\mathbf{Q}_7'$,$\mathbf{Q}_8'$,$\mathbf{Q}_9'$,$\mathbf{Q}_{10}'$},
		x tick label style={rotate=-15,anchor=north},
		ylabel={Runtime (msec)},
		xlabel={},
		ymin=0,ymax=100000,
		ymajorgrids=true,
		xlabel style={below=0.1cm},
		enlarge x limits=0.05,
		legend style={at={(0.65, -0.21)}, 
			legend columns=-1, 
			ylabel near ticks,
			ylabel style={right=0.3cm},
		},
		ybar,
		]
		\addplot
		coordinates {
			($\mathbf{Q}_1'$,7015)($\mathbf{Q}_2'$,1149)($\mathbf{Q}_3'$,51270)($\mathbf{Q}_4'$,1170)($\mathbf{Q}_5'$,1120)($\mathbf{Q}_6'$,138.5)($\mathbf{Q}_7'$,192.3)($\mathbf{Q}_8'$,718.3)($\mathbf{Q}_9'$,807.0)($\mathbf{Q}_{10}'$,2718.3)
		};
		
		\addplot
		coordinates {
			($\mathbf{Q}_1'$,7158)($\mathbf{Q}_2'$,1205)($\mathbf{Q}_3'$,53668)($\mathbf{Q}_4'$,1212)($\mathbf{Q}_5'$,1145)($\mathbf{Q}_6'$,147)($\mathbf{Q}_7'$,220)($\mathbf{Q}_8'$,747)($\mathbf{Q}_9'$,833)($\mathbf{Q}_{10}'$,2763)
		};
		
		\addplot
		coordinates {
			($\mathbf{Q}_1'$,7498)($\mathbf{Q}_2'$,1407)($\mathbf{Q}_3'$,56470)($\mathbf{Q}_4'$,1301)($\mathbf{Q}_5'$,1192)($\mathbf{Q}_6'$,245)($\mathbf{Q}_7'$,253)($\mathbf{Q}_8'$,819)($\mathbf{Q}_9'$,883)($\mathbf{Q}_{10}'$,2845)
		};
		
		\legend{TransE, TransH, TransR}
		
		\end{semilogyaxis}
		\end{tikzpicture}
	}
	\vspace*{-0.1cm}
	\caption{The effect of embedding model on query time }\label{fig:b1}
\end{figure*}
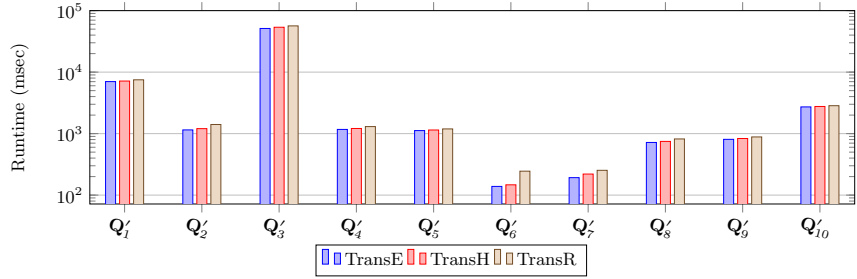

\paragraph{\textbf{Experiment 2.3}} We test the scalability of \texttt{TrQuery} with respect to embedding models. Due to space constraints, we cannot describe in detail results on every dataset and embedding model. We apply \texttt{TransE}, \texttt{TransH}, and \texttt{TransR} in \texttt{TrQuery} on DBpedia with queries $Q_1'$-$Q_{10}'$. Fig.~\ref{fig:b1} reflects the more complex the embedding model is, the longer the query takes, and the impact on runtime is not great.

\subsubsection{\textbf{Experiment 3. Comparison with  state-of-the-art system SAPPER}}
SAPPER~\cite{sapper} is a representative system where some approximate solutions can be provided if a query evaluated no any exact solution. However, SAPPER is unable to recommend approximate solutions, i.e., scoring approximate solutions which are important to users. Moreover, different from our approach based on embedding, SAPPER is based on subgraph indexing which causes inefficiency and connectivity-dependency. In addition, SAPPER only supports edge deletion operator. In this sense, the approximation of SAPPER is in syntax not semantics while our \texttt{TrQuery} can provide the semantic approximation. 

In this experiment, we further compared the efficiency of \texttt{TrQuery} with SAPPER in benchmark dataset. Results overall show that \texttt{TrQuery} is superior to SAPPER in efficiency. Since both the DBpedia and YAGO are disconnected, the comparative experiment was done on a connected subgraph extracted from DBpedia, namely DBpedia*, which contains about thousandth of triples in DBpedia. 
In addition, SAPPER cannot support the query graph without closed path since such query graph will change into a disconnected graph by removing any edge. We designed 6 queries, namely $Q_1^*$-$Q_6^*$, which meet the requirements of SAPPER. 

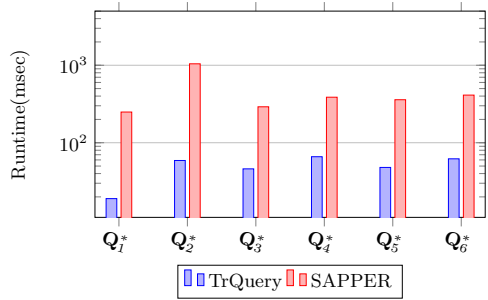
\begin{figure}[!tbp]
	\centering
	\scalebox{0.8}{
		\begin{tikzpicture}
		\begin{semilogyaxis}[
		width=8cm,height=5cm,
		bar width=5pt,
		symbolic x coords={$\mathbf{Q}_1^*$,$\mathbf{Q}_2^*$,$\mathbf{Q}_3^*$,$\mathbf{Q}_4^*$,$\mathbf{Q}_5^*$,$\mathbf{Q}_6^*$},
		x tick label style={rotate=-15,anchor=north},
		ylabel={Runtime(msec)},
		xlabel={},
		ymin=0,ymax=5000,
		ymajorgrids=true,
		xlabel style={below=0.1cm},
		enlarge x limits=0.07,
		legend style={at={(0.5, -0.23)}, 
			anchor=north,
			legend columns=-1, 
			ylabel near ticks,
			ylabel style={right=0.3cm},
		},
		ybar,
		]
		\addplot
		coordinates {
			($\mathbf{Q}_1^*$,19)($\mathbf{Q}_2^*$,59)($\mathbf{Q}_3^*$,46)($\mathbf{Q}_4^*$,66)($\mathbf{Q}_5^*$,48)($\mathbf{Q}_6^*$,62)
		};
		
		\addplot
		coordinates {
			($\mathbf{Q}_1^*$,249)($\mathbf{Q}_2^*$,1044)($\mathbf{Q}_3^*$,291)($\mathbf{Q}_4^*$,387)($\mathbf{Q}_5^*$,359)($\mathbf{Q}_6^*$,412)
		};

		\legend{TrQuery, SAPPER}
		
		\end{semilogyaxis}
		\end{tikzpicture}
	}
	\vspace*{-0.2cm}
	\caption{Comparison of TrQuery and SAPPER in runtime}\label{fig:sapper}
\end{figure}

The results show that SAPPER spent a lot of time in the indexing stage ($>10^6$ \textit{msec}s). Without considering the indexing time of SAPPER, the comparison result of query time is shown in  Fig.~\ref{fig:sapper},  which indicates that \texttt{TrQuery} is much more efficient than SAPPER on each query. On the other hand, the number of approximate solutions returned is comparable with \texttt{TrQuery}. However, SAPPER returns an unsorted solution set, which is unfriendly to users because users prefer to get the answers closest to the correct solution. 

\subsection{Effectiveness Evaluation}

In this part, we evaluate the effectiveness of \texttt{TrQuery}, that is the plausibility of the recomanndation solution set. 
\subsubsection{\textbf{Reciprocal rank}} The first measure we used is the \emph{reciprocal rank} (RR). For a query, RR is the ratio between 1 and the rank at which the first correct answer in the returned result set; or 0 if no correct answer is returned. In any dataset, for all queries,  \texttt{TrQuery} obtained RR = 1, which verifies the correctness of Proposition~\ref{p1}. 

\subsubsection{\textbf{Mean rank}} Another metrics to evaluate the effectiveness of \texttt{TrQuery} is \textit{mean rank} (MR). For a query $Q$, MR is the average of the rank at which each exact solution in the returned result set. Due to Proposition~\ref{p1}, the MR of \texttt{TrQuery} also equals to 1.0 for each query. 

To better evaluate the score function, we firstly destroy the original RDF dataset by deleting some facts, which will cause the query $Q$ to have no exact solution. Then the incomplete data is applied to get a ranked approximate solutions set $S$ for the given query $Q$, and finally calculate the MR value. The better the rankings of correct answers are, the smaller the MR value is. MR=1.0 indicates that all the correct answers are ranked first. The results on DBpedia are shown in the Table~\ref{tab:mr}, which show that \texttt{TrQuery} has a reasonable ordering for approximate solutions. In addition, \texttt{TrQuery-H} and \texttt{TrQuery-R} performs better than \texttt{TrQuery-E} in the evaluation of effectiveness.

\begin{table*}[!tbp]
	\centering
	\caption{Mean rank for $Q_{1}$-$Q_{10}$}\label{tab:mr}
	\vspace*{-0.2cm}
	\scalebox{0.9}{
		\begin{tabular}{|p{0.8cm}|p{1.8cm}|p{0.8cm}|p{0.8cm}|p{0.8cm}|p{0.8cm}|p{0.8cm}|p{0.8cm}|p{0.8cm}|p{0.8cm}|p{0.8cm}|p{0.8cm}|}
			\hline
			\multicolumn{2}{|c|}{Query} & 1 & 2 & 3 & 4 & 5 & 6 & 7 & 8 & 9 & 10\\
			\hline 
			
			\multirow{4}*{MR} & TrQuery\_E & 229.9 & 154.5 & 381.0 & 92.5 & 1.0 & 5.5 & 1.0 & 15.3 & 16.1 & 17.0 \\
			\cline{2-12}
			& TrQuery\_H & 241.3 & 1.5 & 368.0 & 79.2 & 7.0 & 7.5 & 1.0 & 2.0 & 2.1 & 5.0 \\
			\cline{2-12}
			& TrQuery\_R & 240.2 & 21.5 & 361.4 & 127.4 & 88.0 & 1.0 & 1.0 &  11.0 & 11.5 & 12.0 \\
			\hline
		\end{tabular}	
	}
\end{table*}
\subsubsection{\textbf{Recommended approximate solutions of $Q_{a}$}} Finally, we use the recommended solutions of the example query $Q_{a}$ in the Section~\ref{sec:introduction} to illustrate the advantages of \texttt{TrQuery}. There is no exact result when executing $Q_{a}$ on DBpedia. The approximate top-8 results obtained by \texttt{TrQuery} are shown in Table~\ref{tab:example_result}. \texttt{TrQuery} returns an approximately correct result set, which can be verified on Wikipedia. For instance, since Wikipedia's introduction of "Carlo Gabriel Nero" is: "\textit{Carlo Gabriel Nero is an Italian-English \emph{screenwriter} and film director}"\footnote{\url{https://en.wikipedia.org/wiki/Carlo_Gabriel_Nero}}, the top four recommendations can be improved to be correct, even though there is no such fact as (\textit{Carlo\_Gabriel\_Nero, type, ScreenWriter}) exists in DBpedia.

\begin{table*}[!tbp]
	\centering
	\caption{Top-8 approximate solutions for $Q_{a}$}\label{tab:example_result}
	\vspace*{-0.2cm}
	\scalebox{0.85}{
		\begin{tabular}{|p{5cm}<{\centering}|p{3cm}<{\centering}|p{3cm}<{\centering}|p{3cm}<{\centering}|}
			\hline
			?film & ?actor1 & ?actor2 & ?child \\
			\hline
			Camelot\_(film) & Vanessa\_Redgrave & France\_Nero & Carlo\_Gabriel\_Nero\\
			\hline
			Breath\_of\_Life\_(film) & Vanessa\_Redgrave & France\_Nero & Carlo\_Gabriel\_Nero\\
			\hline
			Dropout\_(1970\_film) & Vanessa\_Redgrave & France\_Nero & Carlo\_Gabriel\_Nero\\
			\hline
			A\_Quiet\_Place\_in\_the\_Country & France\_Nero  & Vanessa\_Redgrave & Carlo\_Gabriel\_Nero\\
			\hline
			The\_Rake's\_Progress\_(film) & Rex\_Harrison & Lilli\_Palmer & Carey\_Harrison \\
			\hline
			The\_Long\_Dark\_Hall &  Rex\_Harrison & Lilli\_Palmer & Carey\_Harrison \\
			\hline
			Mission:Impossible\_vs.\_the\_Mob & Barbara\_Bain & Martin\_Landau & Juliet\_Landau \\
			\hline
			Bruce\_Lee:A\_Warrior's\_Journey & Bruce\_Lee & Linda\_Lee\_Cadwell & Shannon\_Lee \\
			\hline
			
		\end{tabular}
	}
\end{table*}

\section{Related Works}\label{sec:related}
In the special case where the edit distance threshold is zero, the problem of graph edit distance becomes subgraph isomorphism, which is NP-complete.
Recently, there are many approaches proposed for approximate subgraph matching~\cite{sapper,tale,path alignment,pbsm,sigma}. TALE~\cite{tale} proposes a novel neighborhood based index (NH-Index) and distinguishes nodes by the importance to the graph structure. In this method, important nodes are matched first and then the match is progressively extended. The method is effective and fast in approximately finding matches in a large graph. SAPPER~\cite{sapper} constructs the hybrid neighborhood unit (HNU) index and takes advantage of pre-generated random spanning trees to accelerate query processing and designs a graph enumeration order to find approximate subgraph matches. 
SIGMA~\cite{sigma} introduces a set-cover based inexact subgraph matching technique and a greedy algorithm to approximate its solution, which takes the identity of the features into account and can distinguish between different features to achieves more filtering power. 
These algorithms use edge misses to measure the quality of a match; and therefore, cannot incorporate the notion of semantics similarity. 
NeMa~\cite{nema} introduces a similarity measure preserving proximity of node pairs and label information. 
However, the structural similarity between query graph and data graph is not considered. Therefore, the computational complexity of NeMa is very large. 
In addition, most of the mentioned works are focused on medical, chemical and protein networks and they are usually not efficient over semantic and social data.

\section{Conclusions}\label{sec:con}

In this paper, we present a novel embedding-based framework \texttt{TrQuery} for approximate query on RDF graphs, which considers both structure and semantic similarity. In this sense, our proposal enriches the current structure-based query recommendation by introducing semantic feature via embedding so that the implicit relationship among queries could be characterized. 
The future work is to improve the efficiency of our \texttt{TrQuery} system. Firstly, the idea of \textit{ranking while matching} can be applied in the future which can stop the execution of the framework as early as possible to improve the overall performance by reducing the redundant verification. Secondly, we will improve efficiency through distributed parallel processing technology.

\section*{Acknowledgments}
This work is supported by the National Key Research and Development Program of China (2017YFC0908401,2016YFB1000603) and the National Natural Science Foundation of China (61672377,61502336).


\begin{thebibliography}{4}
	
	\bibitem{rdf}
	Cyganiak R,  Wood D, and Lanthaler M.
	\newblock {RDF}~1.1 concepts and abstract syntax.
	\newblock {\em W3C recommendation}, 2014. 
	
	\bibitem{sparql}
	Prud'hommeaux E. and Seaborne A. 
	\newblock {SPARQL} query language for {RDF}.
	\newblock W3C Recommendation, 2008.
	
	
	\bibitem{sparql1.1}
	Harris S. and Seaborne A.
	\newblock {SPARQL} 1.1 query language.
	\newblock W3C Recommendation, 2013.
	
	
	\bibitem{perez_sparql_tods}
	P\'erez J,  Arenas M, and Gutierrez C. 
	\newblock Semantics and complexity of {SPARQL}.
	\newblock  {\em ACM Trans. Database Syst.}, 2009, 34(3):article 16.
	
	\bibitem{incomplete_data}
	Sprinzak E, Sattath S, and Margalit H.
	\newblock How reliable are experimental protein-protein in interaction data?
	\newblock {\em J. Molecular Biology}, 2003, 327(5): 919--923.
	
	\bibitem{DBpedia}
	Lehmann J, Isele R, Jakob M, et al. 
	\newblock DBpedia: A large-scale, multilingual knowledge base extracted from Wikipedia.
	\newblock {\em J. Semantic Web}, 2015, 6(2): 167-195.
	
	\bibitem{sapper}
	Zhang S, Yang J, and Jin W. 
	\newblock SAPPER: Subgraph indexing and approximate matching in large graphs.
	\newblock {\em  PVLDB}, 2010, 3(1-2): 1185--1194.
	
	
	
	\bibitem{path alignment}
	De Virgilio R, Maccioni A, and Torlone R. 
	\newblock Approximate querying of RDF graphs via path alignment.
	\newblock {\em  J. Parallel Distrib. Comput.}, 2015, 33(4): 555--581.
	
	
	\bibitem{tale}
	Tian Y and Patel J M. 
	\newblock Tale: A tool for approximate large graph matching,
	\newblock {\em Proc. of ICDE}, 2008: 963--972.
	
	\bibitem{pbsm}
	Chen W, Liu J, Chen Z, Tang X, and Li K.
	\newblock PBSM: An efficient Top-K subgraph matching algorithm.
	\newblock {\em IJPRAI}, 2018, 32(6).
	
	\bibitem{distance1}
	Bunke H and Shearer K. 
	\newblock A graph distance metric based on the maximal common subgraph.
	\newblock {\em Pattern Recogn. lett.}, 1998, 19(3-4): 255--259.
	
	\bibitem{distance2}
	Fernández M L and Valiente G. 
	\newblock A graph distance metric combining maximum common subgraph and minimum common supergraph.
	\newblock {\em Pattern Recogn. lett.}, 2001, 22(6-7): 753--758.
	
	\bibitem{distance3}
	Raymond J W, Gardiner E J, and Willett P. 
	\newblock Rascal: Calculation of graph similarity using maximum common edge subgraphs.
	\newblock {\em Computer J.}, 2002, 45(6): 631--644.
	
	\bibitem{sigma}
	Mongiovi M, Di Natale R, Giugno R, et al. 
	\newblock Sigma: A set-cover-based inexact graph matching algorithm.
	\newblock {\em J. Journal of bioinformatics and computational biology}, 2010, 8(02): 199--218.
	
	\bibitem{nema}
	Khan A, Wu Y, Aggarwal C C,  and Yan X.
	\newblock Nema: Fast graph search with label similarity.
	\newblock {\em PVLDB}, 2013, 6(3): 181--192.
	
	\bibitem{survey of GED}
	Gao X, Xiao B, and Tao D. 
	\newblock A survey of graph edit distance,
	\newblock {\em Pattern Anal. Appl.}, 2010, 13(1): 113--129.
	
	\bibitem{edit distance}
	Riesen K and Bunke H. 
	\newblock Approximate graph edit distance computation by means of bipartite graph matching,
	\newblock {\em Image Vision Comput.}, 2009, 27(7): 950--959.
	
	
	\bibitem{TransE}
	A. Bordes,  N. Usunier,  J. Weston,  and O. Yakhnenko.
	\newblock Translating embeddings for modeling multi-relational data.
	\newblock {\em Proc. of NIPS}, 2013, pp. 2787--2795.
	
	\bibitem{TransH}
	Z. Wang,  J. Zhang,  J. Feng, and  AZ. Chen.
	\newblock Knowledge graph embedding by translating on hyperplanes.
	\newblock {\em Proc. of AAAI}, 2014, pp. 1112--1119.
	
	\bibitem{TransR}
	Y. Lin,  Z. Liu,  X. Zhu,  X. Zhu, and X. Zhu.
	\newblock Learning entity and relation embeddings for knowledge graph completion.
	\newblock {\em Proc. of AAAI'15}, 2015, pp. 2181--2187.
	
	\bibitem{rescal}
	M. Nickel.
	\newblock Tensor factorization for relational learning,
	\newblock {\em Ludwig-Maximilians-Universität München}.
	
	\bibitem{distmult}
	B. Yang,  W. Yih,  X. He,  J. Gao,  and L. Deng.
	\newblock Embedding entities and relations for learning and inference in knowledge bases,
	\newblock {\em arXiv}, 2014.
	
	\bibitem{hole}
	M. Nickel,  L. Rosasco,  T. Poggio.
	\newblock Holographic Embeddings of Knowledge Graphs
	\newblock {\em AAAI}, 2016: 1955-1961.
	
	
	\bibitem{TransD}
	G. Ji,  S. He,  L. Xu,  K. Liu,  J. Zhao.
	\newblock Knowledge Graph Embedding via Dynamic Mapping Matrix.
	\newblock {\em ACL}, (1). 2015: 687--696.
	
	\bibitem{yago2}
	J. Hoffart, F.M. Suchanek, K. Berberich, and G. Weikum.
	\newblock {YAGO2}: A spatially and temporally enhanced knowledge base from Wikipedia.
	\newblock {\em Artif. Intell.}, 2013, 194: 28--61.
	
	\bibitem{representation review}
	M. Nickel, K. Murphy, V. Tresp, E. Gabrilovich. (2016). 
	\newblock A review of relational machine learning for knowledge graphs. 
	\newblock {\em J. Proceedings of the IEEE}, 104(1): 11--33.
	
	\bibitem{embedding_survey}
	Q. Wang,  Z. Mao, B. Wang, L. Guo. (2017).
	\newblock Knowledge graph embedding: A survey of approaches and applications.
	\newblock {\em  J. IEEE Trans. Knowl. Data Eng.}, 29(12):2724--2743.
	
	
	
	
	
	
	
	
	
	
	
\end{thebibliography}
\end{document}